\documentclass{ieeeaccess}
\usepackage{cite}
\usepackage{amsmath,amssymb,amsfonts}
\usepackage{algorithmic}
\usepackage{graphicx}
\usepackage{textcomp}

\usepackage{balance}
\usepackage{multirow}
\usepackage[table,xcdraw]{xcolor}
\usepackage{verbatim}

\usepackage[colorlinks,urlcolor=blue,linkcolor=blue,citecolor=blue]{hyperref}

\def\BibTeX{{\rm B\kern-.05em{\sc i\kern-.025em b}\kern-.08em
    T\kern-.1667em\lower.7ex\hbox{E}\kern-.125emX}}
\begin{document}
\history{Submitted to IEEE Access}
\doi{10.1109/ACCESS.2017.DOI}

\newtheorem{theorem}{Theorem}
\newtheorem{lemma}{Lemma}
\newcommand{\vulname}{SelfLogAttack}

\title{My SIM is Leaking My Data: \\Exposing Self-Login Privacy Breaches in Smartphones}

\author{
\uppercase{A. Coletta}\authorrefmark{1} \IEEEmembership{Student Member, IEEE},
\uppercase{G. Maselli}\authorrefmark{1}\IEEEmembership{Member, IEEE},
\uppercase{M. Piva}\authorrefmark{1}\IEEEmembership{Student Member, IEEE},
\uppercase{D. Silvestri}\authorrefmark{1}, 
\uppercase{and F. Restuccia}\authorrefmark{2}\IEEEmembership{Member, IEEE}
}

\address[1]{Sapienza, University of Rome, Rome, Italy, (e-mail: \{coletta, maselli, piva, silvestri.d\}@di.uniroma1.it)}

\address[2]{Northeastern University, Boston, USA (e-mail: frestuc@northeastern.edu)}




\begin{abstract}
In the latest years the attention on management of users' personal data has increased significantly to ensure security and privacy. Several new regulations and public entities are born to guarantee and regulate protection of user data. Nevertheless, users are still exposed to a high number of issues and leaks. 
In this paper we expose a new security leak for smartphone users, which allows to stole user personal data by accessing the mobile operator user page when auto-login is employed. We show how any "apparently" genuine app can steal these data from some mobile operators; or how an attacker can stole them by exploiting a shared internet connection, e.g., through hot-spot granted by the user.
We analyse different mobile operator companies to demonstrate the highlighted issues, and we discover that more than 40 millions of mobile smartphones are vulnerable. Finally, we propose some possible countermeasures.
\end{abstract}


\titlepgskip=-15pt

\maketitle

\section{Introduction}

Smartphones have become \textit{the} most popular mobile device, especially thanks to their fast evolution and their capabilities. They have become an invaluable tool for everyday life and are used for any kind of activities, from simple video-call, photos, or email, to social media interaction and sport tracking activities. 

On the other hand, the pervasiveness of smartphones, together with a myriad of mobile applications, has highly increased the related security and privacy threats for user data \cite{ali_rah,frest-iot}. As smartphones contain a trove of private information, users are continuously exposed to privacy issues. Mobile phones own a plethora of details about our personal life: locations, messages, calls. That makes smartphone privacy issues more important than ever -- personal life details are private and must remain so. To guarantee and protect user data, over the last few years several public regulations and entities were born, such as GDPR \cite{ref0} in Europe and CCPA \cite{ref_ccpa} in the USA. These regulations set best practices and rules to manage, store and protect the collected user data. Nevertheless, users are still exposed to several security issues and privacy leaks, and most of them are only partially known. 

In this paper, we aim to highlight a severe privacy vulnerability of mobile operators (\textit{i.e.}, the telecommunication companies) which threatens to expose users' data to attacks at the expense of a faster and frictionless user experience. In particular, mobile operators (MOs) often provide a dedicated web page or an application for their users, to check their billing accounts. With the advent of different commercial offers, customers can be charged depending on many factors, and to keep them aware of their expenses, the mobile operators must provide access to their accounts. 

The vulnerability works as follows: mobile operators often provide a user-friendly service to access auto authenticated web pages or mobile apps. When a user is surfing the Internet through her mobile connection she can automatically access her billing account, and her personal information, the amount of credit available, the number of calls done, the amount of gigabytes consumed, the text messages delivered, and so on. She can also activate or disable services, without any explicit authentication. 


\Figure[t!](topskip=0pt, botskip=0pt, midskip=0pt)[width=0.5\linewidth]{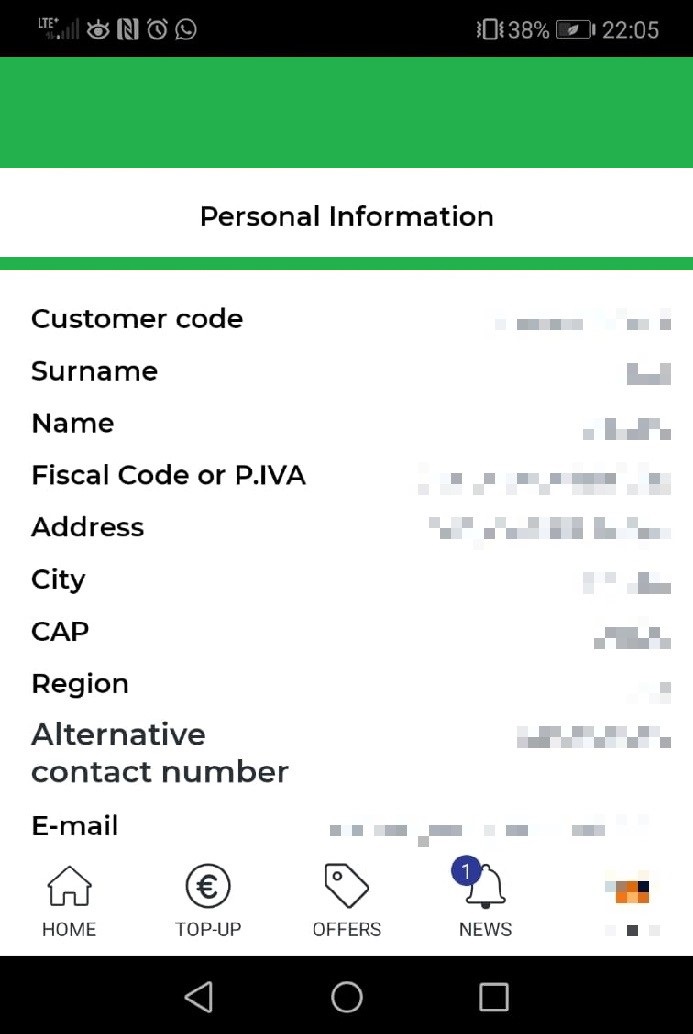}
{Example of data that can be stolen by a malicious app installed on a user smartphone. The app is able to access all the user personal details: name and surname; customer identification code; personal identification numbers (i.e., Fiscal code or P.IVA); address data (i.e., street address, city, region and state); and contact info, such email and alternative number.\label{fig:mo1}}

These services often use the "Frictionless" design pattern \cite{frictionless} -- in a nutshell, users should be able to access services with the lowest number of obstacles possible. One famous example is the "one-click buying" service developed by Amazon. Once that the user has registered its delivery address and its payment methods, it has to do only one click to buy a product. Without asking for logins or double confirmation, the user can buy in a fast and obstacle free way. 

While this system makes life easier for the users, it comes with several threats to privacy. For example, with the Amazon service, as soon as an order is created an email is sent to the user performing the action. If an unauthorized user is using the "one-click buying" service, the owner of the account is immediately informed and can interrupt the operation, since there is a time between the order creation and the effective delivery.

\begin{figure}[t]
\centering
\includegraphics[width=\linewidth]{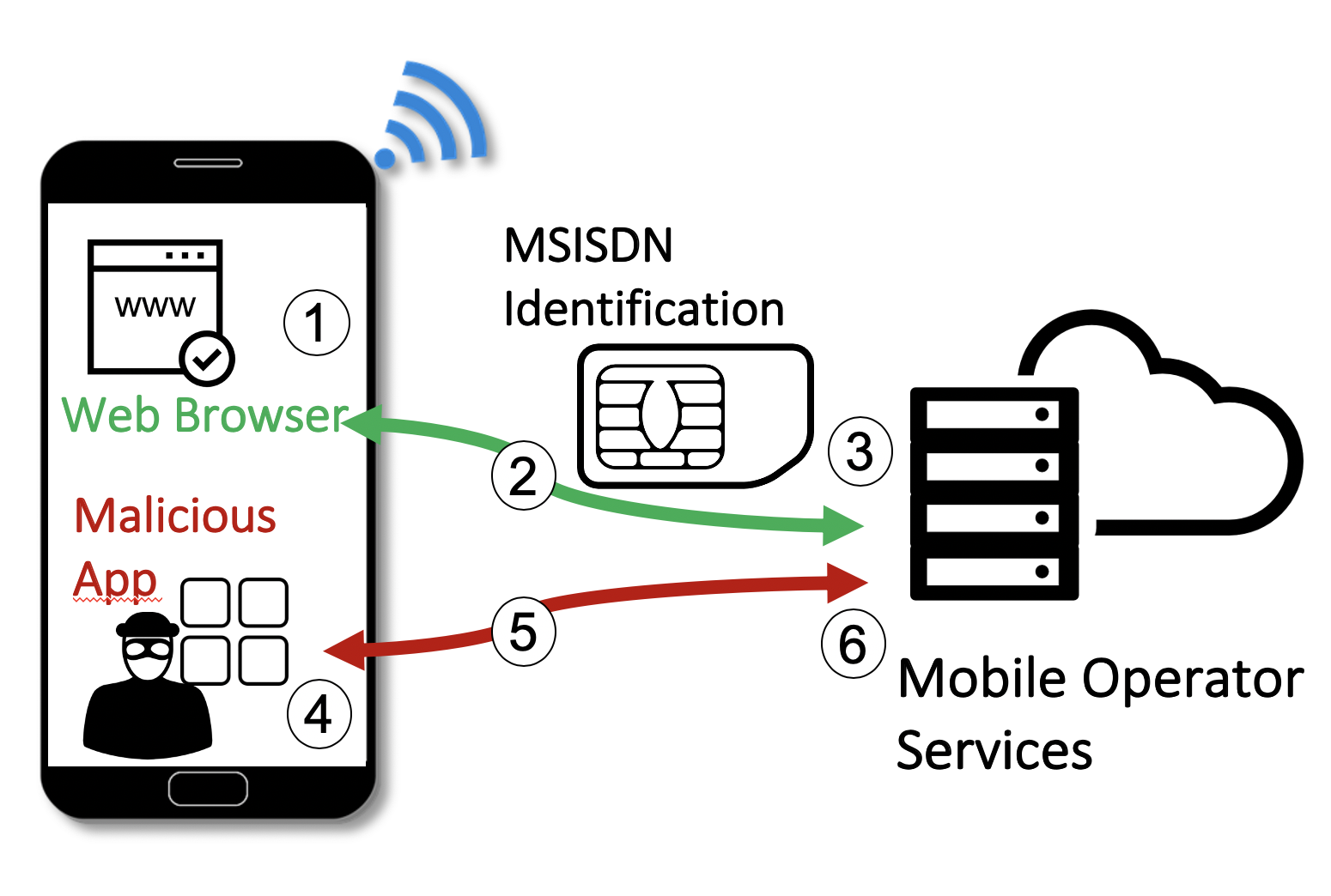}
\caption{
The figure shows how any (malicious) mobile app can steal the user data from the mobile operator servers. The method takes advantage of an "apparently" genuine mobile application. The app, through simple HTTP requests, can access user data and services from the mobile operator servers, it collects them, and eventually it sends these data to some malicious server. 
}
\label{fig:img1}
\end{figure}

\begin{figure}[t]
\centering
\includegraphics[width=\linewidth]{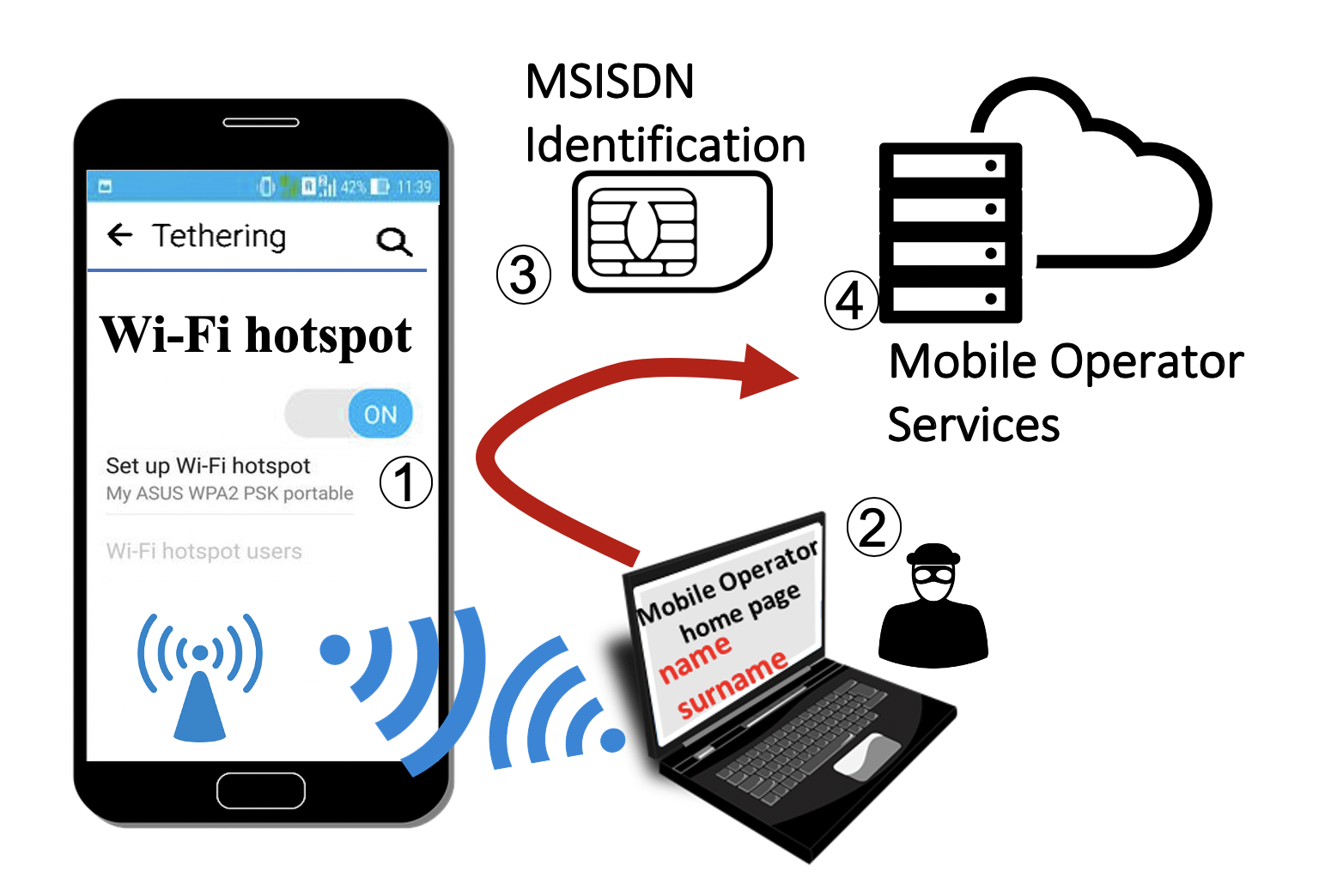}
\caption{The figure shows how some user private data may be stolen by a malicious attacker through a shared hot-spot connection.
}
\label{fig:img2}
\end{figure}

Following the same approach, many telecommunication companies allow their mobile users to access some services without any login, as the company is able to identify from which SIM, subscriber identity module, the traffic is generated. The frictionless service improves the user experience, but it may also expose private information. For example, all the applications installed on the smartphone may access the user name, address, id code, or phone number. Figure \ref{fig:mo1} shows a real example of the data that may be stolen by a mobile application. Moreover, if the user grants an attacker to use his Internet connection (\textit{e.g.}, through a hotspot service), the latter can access his billing account. In general, without an explicit authentication method, any application, or browser, can access the user data on the mobile operator server. Figure \ref{fig:img1} and Figure \ref{fig:img2} explain the two attacks.


The attack depicted in Figure \ref{fig:img1} can be accomplished by any application installed on a smartphone. A number of mobile operators allow their users to access personal information through the web browser (step 1 of the figure), without requiring any password. The authentication is done (step 2) through the identification of the MSISDN which is generating the traffic. Mobile Operator services uniquely identify the user and return the required information (step 3). Similarly, an app installed on the user smartphone may behave like the web browser (step 4). Its requests would be authenticated (step 5) exactly like standard requests (as in step 2) and the Mobile Operator Service would return the requested data. 

The attacker may also exploit the hotspot functionality, as depicted in Figure \ref{fig:img2}. Specifically, if the attacker is connected to the Internet through a user mobile hotspot (step 1) and sends any request on the Internet, then Mobile Operator services will identify those requests as generated by the user MSISDN (step 3), and hence as legal (authentic) requests.


In this paper, we investigate the aforementioned privacy issues and possible solutions. In particular, we describe the mobile operator vulnerabilities, we test seven different Italian mobile operators to explore the different possibilities to access user data; we propose some possible countermeasures; we discuss some related work and finally draw conclusions. 

%
%

\section{Threat Analysis}\label{sec:description}

The key concept behind the threat is that self-authentication allows identified users to easily access a number of services without any explicit login. Mobile operators, in particular, leverage this mechanism to speed up the access time for their users. Moreover, users can access information on its contract with the mobile operator without any explicit login. The mechanism hinges on the fact that mobile operators identify from which SIM the traffic is generated and, as the SIM is associated with a unique person, they can easily identify the owner of the SIM. This information includes, among others, personal data (\textit{e.g.}, name, surname, or address) and/or data about its mobile contract, such as the SIM security code, the active commercial offers and so on.

Furthermore, users are able to perform different operations or to access several services. They can activate offers or transfer credit to other users, without any login. When the smartphone is effectively protected, for example with a passcode or with biometric controls, an attacker cannot use the user's smartphone to access these auto-logged pages. 
However, these pages may be available to all the applications installed on the user smartphone, just at an HTTP request away. 

In addition, access to the Internet is nowadays automatically allowed to any app both on Android and iOS, without any specific permission\cite{android-permission}\cite{ios-permission}. This means that any app could be able to access the mobile operator pages, do the self-login and extract information on the SIM owner. The same app may also be able to perform operations on those pages, such as activating paid services or downloading the user voice mail.

While the user in case of paid services can recognize the unauthorized access, in case of data extraction (\textit{e.g.}, name, address, call list), she is agnostic about the fraudulent access.  Even if almost all the mobile applications are distributed by authorized stores (\textit{e.g.}, Google Play store for Android), they can still be harmful. These checks are performed both automatically and by humans. However, as the attack requires only HTTP requests, they are not detected as malware or fraudulent. Potentially, all the applications on iOS and Android stores may perform this attack,  without the user knowing that. In other words, depending on the mobile operator, an app on smartphones may perfectly access data, and extract user information, without being detected.

Moreover, a similar attack can be performed by exploiting the WiFi hotspot functionality. Indeed, when a smartphone acts as a hotspot, the self-authentication feature is extended to all the devices that are connected to it. Therefore, even without installing any malicious app on the smartphone someone could execute an attack simply by connecting to the hotspot.  

We performed an analysis of the major mobile operators in Italy, referring to the publicly accessible number of customers they have. We discovered that the most important operators have weak protections against this attack, leaving more than 80\% of mobile smartphone exposed.
%
%
On the other hand, smaller companies implemented strong protections against this attack. For the companies that we analyzed we present the protections that we found, and report if we were able to overcome them, in addition to which information we were able to extract and which active operations were allowed to do.

%
%
\section{List of Vulnerabilities}\label{sec:experiments}

Confidentiality, Integrity, and Availability are the three key principles which should be guaranteed in any secure program. Even two of these principles --- confidentiality and integrity --- can be violated by exploiting {\vulname}. In the following, we present the attacks we identified, classifying them into two groups: passive attacks (\textit{i.e.} they only extract user's information without modifying anything), and active attacks (\textit{i.e.} (they actively perform operations by modifying the user services or credit).
We tested seven major Italian mobile operators against these vulnerabilities (we call them Mobile Operator 1 or MO1, Mobile Operator 2, etc.). For each operator we explored the possibility to perform self-login through a browser connection or through the operator mobile application. In the case of self-login capable pages, we tried to perform programmatic login and to interact with the operator service. Table \ref{tab:my-table} summarizes the information we were able to extract and the operations that we were able to perform.

\begin{table*}[ht]
\centering
\caption{Vulnerabilities of analyzed Mobile Operators}
\label{tab:my-table}
\begin{tabular}{lllllllll}
\hline
\multicolumn{3}{|c|}{} &
  \multicolumn{6}{c|}{Mobile Operator} \\ \cline{4-9} 
\multicolumn{3}{|c|}{\multirow{-2}{*}{Vulnerabilities}} &
  \multicolumn{1}{l|}{MO 1} &
  \multicolumn{1}{l|}{MO 2} &
  \multicolumn{1}{l|}{MO 3} &
  \multicolumn{1}{l|}{MO 4} &
  \multicolumn{1}{l|}{MO 5} &
  \multicolumn{1}{l|}{MO 6 and MO 7} \\ \hline
\multicolumn{1}{|l|}{} &
  \multicolumn{1}{l|}{} &
  \multicolumn{1}{l|}{Name} &
  \multicolumn{1}{l|}{{\color[HTML]{FE0000} \textbf{Y}}} &
  \multicolumn{1}{l|}{{\color[HTML]{FE0000} \textbf{Y}}} &
  \multicolumn{1}{l|}{{\color[HTML]{FE0000} \textbf{Y}}} &
  \multicolumn{1}{l|}{{\color[HTML]{FE0000} \textbf{Y}}} &
  \multicolumn{1}{l|}{{\color[HTML]{32CB00} \textbf{N}}} &
  \multicolumn{1}{l|}{{\color[HTML]{32CB00} }} \\ \cline{3-8}
\multicolumn{1}{|l|}{} &
  \multicolumn{1}{l|}{} &
  \multicolumn{1}{l|}{Surname} &
  \multicolumn{1}{l|}{{\color[HTML]{FE0000} \textbf{Y}}} &
  \multicolumn{1}{l|}{{\color[HTML]{FE0000} \textbf{Y}}} &
  \multicolumn{1}{l|}{{\color[HTML]{FE0000} \textbf{Y}}} &
  \multicolumn{1}{l|}{{\color[HTML]{FE0000} \textbf{Y}}} &
  \multicolumn{1}{l|}{{\color[HTML]{32CB00} \textbf{N}}} &
  \multicolumn{1}{l|}{{\color[HTML]{32CB00} }} \\ \cline{3-8}
\multicolumn{1}{|l|}{} &
  \multicolumn{1}{l|}{} &
  \multicolumn{1}{l|}{Mobile Number (MSISDN)} &
  \multicolumn{1}{l|}{{\color[HTML]{FE0000} \textbf{Y}}} &
  \multicolumn{1}{l|}{{\color[HTML]{FE0000} \textbf{Y}}} &
  \multicolumn{1}{l|}{{\color[HTML]{FE0000} \textbf{Y}}} &
  \multicolumn{1}{l|}{{\color[HTML]{FE0000} \textbf{Y}}} &
  \multicolumn{1}{l|}{{\color[HTML]{FE0000} \textbf{Y}}} &
  \multicolumn{1}{l|}{{\color[HTML]{32CB00} }} \\ \cline{3-8}
\multicolumn{1}{|l|}{} &
  \multicolumn{1}{l|}{} &
  \multicolumn{1}{l|}{Tax Code} &
  \multicolumn{1}{l|}{{\color[HTML]{FE0000} \textbf{N*}}} &
  \multicolumn{1}{l|}{{\color[HTML]{FE0000} \textbf{Y}}} &
  \multicolumn{1}{l|}{{\color[HTML]{FE0000} \textbf{Y}}} &
  \multicolumn{1}{l|}{{\color[HTML]{32CB00} \textbf{N}}} &
  \multicolumn{1}{l|}{{\color[HTML]{32CB00} N}} &
  \multicolumn{1}{l|}{{\color[HTML]{32CB00} }} \\ \cline{3-8}
\multicolumn{1}{|l|}{} &
  \multicolumn{1}{l|}{} &
  \multicolumn{1}{l|}{Birth Date} &
  \multicolumn{1}{l|}{{\color[HTML]{FE0000} \textbf{Y}}} &
  \multicolumn{1}{l|}{{\color[HTML]{FE0000} \textbf{Y}}} &
  \multicolumn{1}{l|}{{\color[HTML]{FE0000} \textbf{Y}}} &
  \multicolumn{1}{l|}{{\color[HTML]{32CB00} \textbf{N}}} &
  \multicolumn{1}{l|}{{\color[HTML]{32CB00} N}} &
  \multicolumn{1}{l|}{{\color[HTML]{32CB00} }} \\ \cline{3-8}
\multicolumn{1}{|l|}{} &
  \multicolumn{1}{l|}{} &
  \multicolumn{1}{l|}{Birth Place} &
  \multicolumn{1}{l|}{{\color[HTML]{FE0000} \textbf{Y}}} &
  \multicolumn{1}{l|}{{\color[HTML]{FE0000} \textbf{Y}}} &
  \multicolumn{1}{l|}{{\color[HTML]{FE0000} \textbf{Y}}} &
  \multicolumn{1}{l|}{{\color[HTML]{32CB00} \textbf{N}}} &
  \multicolumn{1}{l|}{{\color[HTML]{32CB00} \textbf{N}}} &
  \multicolumn{1}{l|}{{\color[HTML]{32CB00} }} \\ \cline{3-8}
\multicolumn{1}{|l|}{} &
  \multicolumn{1}{l|}{\multirow{-7}{*}{Personal Data}} &
  \multicolumn{1}{l|}{Address of Residence} &
  \multicolumn{1}{l|}{{\color[HTML]{FE0000} \textbf{Y}}} &
  \multicolumn{1}{l|}{{\color[HTML]{FE0000} \textbf{Y}}} &
  \multicolumn{1}{l|}{{\color[HTML]{FE0000} \textbf{Y}}} &
  \multicolumn{1}{l|}{{\color[HTML]{FE0000} \textbf{Y}}} &
  \multicolumn{1}{l|}{{\color[HTML]{32CB00} N}} &
  \multicolumn{1}{l|}{{\color[HTML]{32CB00} }} \\ \cline{2-8}
\multicolumn{1}{|l|}{} &
  \multicolumn{1}{l|}{} &
  \multicolumn{1}{l|}{Active Offers} &
  \multicolumn{1}{l|}{{\color[HTML]{FE0000} \textbf{Y}}} &
  \multicolumn{1}{l|}{{\color[HTML]{FE0000} \textbf{Y}}} &
  \multicolumn{1}{l|}{{\color[HTML]{FE0000} \textbf{Y}}} &
  \multicolumn{1}{l|}{{\color[HTML]{FE0000} \textbf{Y}}} &
  \multicolumn{1}{l|}{{\color[HTML]{FE0000} \textbf{Y}}} &
  \multicolumn{1}{l|}{{\color[HTML]{32CB00} }} \\ \cline{3-8}
\multicolumn{1}{|l|}{} &
  \multicolumn{1}{l|}{} &
  \multicolumn{1}{l|}{Credit} &
  \multicolumn{1}{l|}{{\color[HTML]{FE0000} \textbf{Y}}} &
  \multicolumn{1}{l|}{{\color[HTML]{FE0000} \textbf{Y}}} &
  \multicolumn{1}{l|}{{\color[HTML]{FE0000} \textbf{Y}}} &
  \multicolumn{1}{l|}{{\color[HTML]{FE0000} \textbf{Y}}} &
  \multicolumn{1}{l|}{{\color[HTML]{FE0000} \textbf{Y}}} &
  \multicolumn{1}{l|}{{\color[HTML]{32CB00} }} \\ \cline{3-8}
\multicolumn{1}{|l|}{} &
  \multicolumn{1}{l|}{} &
  \multicolumn{1}{l|}{PIN} &
  \multicolumn{1}{l|}{{\color[HTML]{FE0000} \textbf{Y}}} &
  \multicolumn{1}{l|}{{\color[HTML]{32CB00} \textbf{N}}} &
  \multicolumn{1}{l|}{{\color[HTML]{32CB00} \textbf{N}}} &
  \multicolumn{1}{l|}{{\color[HTML]{32CB00} \textbf{N}}} &
  \multicolumn{1}{l|}{{\color[HTML]{32CB00} N}} &
  \multicolumn{1}{l|}{{\color[HTML]{32CB00} }} \\ \cline{3-8}
\multicolumn{1}{|l|}{} &
  \multicolumn{1}{l|}{\multirow{-4}{*}{SIM Data}} &
  \multicolumn{1}{l|}{PUK} &
  \multicolumn{1}{l|}{{\color[HTML]{FE0000} \textbf{Y}}} &
  \multicolumn{1}{l|}{{\color[HTML]{32CB00} \textbf{N}}} &
  \multicolumn{1}{l|}{{\color[HTML]{32CB00} \textbf{N}}} &
  \multicolumn{1}{l|}{{\color[HTML]{FE0000} \textbf{Y***}}} &
  \multicolumn{1}{l|}{{\color[HTML]{FE0000} \textbf{Y}}} &
  \multicolumn{1}{l|}{{\color[HTML]{32CB00} }} \\ \cline{2-8}
\multicolumn{1}{|l|}{} &
  \multicolumn{1}{l|}{} &
  \multicolumn{1}{l|}{Calls} &
  \multicolumn{1}{l|}{{\color[HTML]{FE0000} \textbf{Y**}}} &
  \multicolumn{1}{l|}{{\color[HTML]{32CB00} \textbf{N}}} &
  \multicolumn{1}{l|}{{\color[HTML]{FE0000} \textbf{Y**}}} &
  \multicolumn{1}{l|}{{\color[HTML]{FE0000} \textbf{Y}}} &
  \multicolumn{1}{l|}{{\color[HTML]{32CB00} N}} &
  \multicolumn{1}{l|}{{\color[HTML]{32CB00} }} \\ \cline{3-8}
\multicolumn{1}{|l|}{} &
  \multicolumn{1}{l|}{} &
  \multicolumn{1}{l|}{SMS Senders} &
  \multicolumn{1}{l|}{{\color[HTML]{FE0000} \textbf{Y**}}} &
  \multicolumn{1}{l|}{{\color[HTML]{32CB00} \textbf{N}}} &
  \multicolumn{1}{l|}{{\color[HTML]{FE0000} \textbf{Y**}}} &
  \multicolumn{1}{l|}{{\color[HTML]{FE0000} \textbf{Y}}} &
  \multicolumn{1}{l|}{{\color[HTML]{32CB00} N}} &
  \multicolumn{1}{l|}{{\color[HTML]{32CB00} }} \\ \cline{3-8}
\multicolumn{1}{|l|}{\multirow{-14}{*}{\begin{tabular}[c]{@{}l@{}}Passive\\  Attack\end{tabular}}} &
  \multicolumn{1}{l|}{\multirow{-3}{*}{\begin{tabular}[c]{@{}l@{}}Historical\\ Data\end{tabular}}} &
  \multicolumn{1}{l|}{Voice Mail} &
  \multicolumn{1}{l|}{{\color[HTML]{32CB00} \textbf{N}}} &
  \multicolumn{1}{l|}{{\color[HTML]{32CB00} \textbf{N}}} &
  \multicolumn{1}{l|}{{\color[HTML]{32CB00} \textbf{N}}} &
  \multicolumn{1}{l|}{{\color[HTML]{FE0000} \textbf{Y}}} &
  \multicolumn{1}{l|}{{\color[HTML]{32CB00} N}} &
  \multicolumn{1}{l|}{{\color[HTML]{32CB00} }} \\ \cline{1-8}
\multicolumn{1}{|l|}{} &
  \multicolumn{1}{l|}{Services} &
  \multicolumn{1}{l|}{Activate Services} &
  \multicolumn{1}{l|}{{\color[HTML]{FE0000} \textbf{Y}}} &
  \multicolumn{1}{l|}{{\color[HTML]{FE0000} \textbf{Y}}} &
  \multicolumn{1}{l|}{{\color[HTML]{32CB00} \textbf{N}}} &
  \multicolumn{1}{l|}{{\color[HTML]{FE0000} \textbf{Y}}} &
  \multicolumn{1}{l|}{{\color[HTML]{32CB00} N}} &
  \multicolumn{1}{l|}{{\color[HTML]{32CB00} }} \\ \cline{2-8}
\multicolumn{1}{|l|}{} &
  \multicolumn{1}{l|}{} &
  \multicolumn{1}{l|}{Change Password} &
  \multicolumn{1}{l|}{{\color[HTML]{FE0000} \textbf{Y}}} &
  \multicolumn{1}{l|}{{\color[HTML]{32CB00} \textbf{N}}} &
  \multicolumn{1}{l|}{{\color[HTML]{32CB00} \textbf{N}}} &
  \multicolumn{1}{l|}{{\color[HTML]{32CB00} \textbf{N}}} &
  \multicolumn{1}{l|}{{\color[HTML]{32CB00} \textbf{N}}} &
  \multicolumn{1}{l|}{{\color[HTML]{32CB00} }} \\ \cline{3-8}
\multicolumn{1}{|l|}{} &
  \multicolumn{1}{l|}{\multirow{-2}{*}{Password}} &
  \multicolumn{1}{l|}{Change PIN} &
  \multicolumn{1}{l|}{{\color[HTML]{32CB00} \textbf{N}}} &
  \multicolumn{1}{l|}{{\color[HTML]{32CB00} \textbf{N}}} &
  \multicolumn{1}{l|}{{\color[HTML]{32CB00} \textbf{N}}} &
  \multicolumn{1}{l|}{{\color[HTML]{32CB00} \textbf{N}}} &
  \multicolumn{1}{l|}{{\color[HTML]{32CB00} \textbf{N}}} &
  \multicolumn{1}{l|}{{\color[HTML]{32CB00} }} \\ \cline{2-8}
\multicolumn{1}{|l|}{\multirow{-4}{*}{\begin{tabular}[c]{@{}l@{}}Active\\ Attack\end{tabular}}} &
  \multicolumn{1}{l|}{Credit} &
  \multicolumn{1}{l|}{Transfer Credit} &
  \multicolumn{1}{l|}{{\color[HTML]{32CB00} \textbf{N}}} &
  \multicolumn{1}{l|}{{\color[HTML]{FE0000} \textbf{Y}}} &
  \multicolumn{1}{l|}{{\color[HTML]{32CB00} \textbf{N}}} &
  \multicolumn{1}{l|}{{\color[HTML]{32CB00} \textbf{N}}} &
  \multicolumn{1}{l|}{{\color[HTML]{32CB00} \textbf{N}}} &
  \multicolumn{1}{l|}{{\color[HTML]{32CB00} }} \\ \cline{1-8}
\multicolumn{3}{|l|}{Self-Login Always Active} &
  \multicolumn{1}{l|}{{\color[HTML]{32CB00} \textbf{N}}} &
  \multicolumn{1}{l|}{{\color[HTML]{FE0000} \textbf{Y}}} &
  \multicolumn{1}{l|}{{\color[HTML]{FE0000} \textbf{Y}}} &
  \multicolumn{1}{l|}{{\color[HTML]{FE0000} \textbf{Y}}} &
  \multicolumn{1}{l|}{{\color[HTML]{FE0000} \textbf{Y}}} &
  \multicolumn{1}{l|}{\multirow{-19}{*}{{\color[HTML]{32CB00} \textbf{\begin{tabular}[c]{@{}l@{}}\centering Password Required\end{tabular}}}}} \\ \hline
\multicolumn{9}{l}{\begin{tabular}[c]{@{}l@{}}\\** Even if not directly available, the Tax Code can be generated from the other information extracted.\\ ** The last three digits of each number are obfuscated.\\ *** In case of disclosure an email is sent to the SIM owner.\end{tabular}}
\end{tabular}
\end{table*}

\subsection{Confidentiality - Passive Attacks}

Through API interaction, we have been able to extract the following information:
\paragraph{Personal Data}
It is possible to extract a wide number of private personal data such as name, surname, mobile number, tax code, birth date, birth place and the address of residence. From three companies, all these data are fully available. Mobile Operator 4 hides information regarding birth date and place, making impossible to generate the tax code. Mobile Operator 5 hides all these data, apart from the Mobile International ISDN Number (MSISDN) which is the user mobile number \cite{itu}. However, even sharing only the mobile number allows attackers to track a user and discover private information. The MSISDN is composed of Country Code, Number Planning Area and Subscriber Number. This allows to discover the user country and operator. Both Mobile Operators 6 and 7 require the user password to access those data.
\paragraph{SIM Data}
All the mobile operators affected by {\vulname} allow unauthorized access to a list of the offers that are active on the user contract. They also show the amount of call, SMS, navigation traffic and available credit. With one mobile operator, MO 1, it is also possible to access the current PIN and PUK of the user SIM. PUK code is also available with MO 4 and MO 5. However, MO 4 sends an email to the user as soon as a request to access the PUK is made.
\paragraph{Historical Data}
Mobile Operator 1 and 3 partially reveal, without authentication, the last called phone numbers and SMS recipients. MO 4 instead fully shows those numbers. MO 4 also allows access to the audio files of the user voicemail. 

\subsection{Integrity - Active Attacks}
We explored the possibility to perform a number of active operations on the user's account:
\paragraph{Activate Services}
MO 3 and 5 allow to activate services, which requires payment, and to disable some \emph{spending limit}, i.e., a form of protection for users from too high billing that disables the user mobile when such a limit is reached.
With MO 4 we have also been able to activate a "User not available" mode, which automatically rejects all incoming calls.
\paragraph{Password} When self-authentication is disabled users are required to insert a username, typically the user phone number, and a password, in order to access the Mobile Operator portal containing user data. MO 1 allows changing the user's personal password, that one used to access the MO portal, without requiring any authentication and/or insertion of the old password.

\paragraph{Managing Credit}~MO 2 allows to transfer credit to other customers of the same Mobile Operator. All the other recharge operations are protected with a password.

\paragraph{Privacy by default} Apart from MO 1, self login functionality cannot be deactivated by customers. MO 1 however has this service active by default, breaking the "Privacy by default" pattern \cite{privacybydefault}.

\subsection{Attack method}
For each Mobile Operator, we give some details on how we have been able to exploit the vulnerability.
\paragraph{Mobile Operator 1}
Exploring the information web page that is accessible with self-login it is possible to find many exposed endpoints, able to receive HTTP requests.
While this company exposes a wide number of users' details, it is the only one that allows customers to disable self-login. However, against the "Privacy by default" pattern, self-login in automatically enabled on new contracts.

\paragraph{Mobile Operator 2}
This operator allows to identify a number of exposed API on the information web page that is accessible with self-login. From those API, through unauthorized HTTP POST requests, it is possible to obtain personal information in different formats, like JSON or XML. Requests have to be executed using TLS, and no controls are made on sent headers.

\paragraph{Mobile Operator 3}
While this company does not offer any self-logged access through browsers, its mobile application allows access to a wide number of functionalities without authentication. This information is exposed through a series of API that can be easily extracted from the mobile application. In this case, it would be possible to collect these endpoints, decompiling the Android application.

\paragraph{Mobile Operator 4}

This company exposes a website where a user, using its mobile connection, can self-login and access much information. The landing page is equipped with a Javascript function which verifies, making a server-side call to an authentication server, if the user is using a mobile connection and if the SIM is owned by the operator. If the HTTP request is successful, the user is authenticated to its personal page; otherwise, she is redirected to another page with a cookie. Making a simple HTTP GET request to the page is not enough to access that page. However, if an attacker makes an HTTP GET request to the authentication server, she is able to obtain an authenticated cookie session. With this cookie set, the attacker is able to extract the user information with a series of HTTP GET requests.

\paragraph{Mobile Operator 5}
This company exposes a web page, always accessible only with the mobile connection, which shows the user mobile number and other information regarding the active offers. However, analyzing the company mobile application, it is possible to find other APIs which are self-authenticated.

%
%

\section{Attack Countermeasures}\label{sec:Countermeasures}

We analyze a number of countermeasures that allow users to have a safer experience without privacy issues. For each proposed solution we analyze its \emph{effectiveness} --- how much it really protects users --- its \emph{strength} --- how complex it is for an attacker to overcome the protection --- and its \emph{impact on the user} --- how much the countermeasure implementation impacts the user experience. For example, a control made with an SMS Authentication is very effective, as all the endpoints are unreachable without the code received in the SMS. However, for an attacker it could be possible to access the SMS in an easy way, for example requiring specific permission to a not careful user, compromising the strength.

\begin{table}[hpt!]
\centering
\caption{Contermeasures}
\label{tab:my-table2}
\resizebox{\linewidth}{!}{
\begin{tabular}{l|l|l|l|}
\cline{2-4}
                                              & Effective                        & Strong                            & Frictionless                       \\ \hline
\multicolumn{1}{|l|}{Headers Control}         & {\color[HTML]{FE0000} \textbf{Low}}    & {\color[HTML]{FE0000} \textbf{Low}}    & {\color[HTML]{009901} \textbf{High}}   \\ \hline
\multicolumn{1}{|l|}{In-App Certificate}     & {\color[HTML]{F8A102} \textbf{Medium}} & {\color[HTML]{F8A102} \textbf{Medium}} & {\color[HTML]{F8A102} \textbf{Medium}} \\ \hline
\multicolumn{1}{|l|}{SMS Authentication}      & {\color[HTML]{009901} \textbf{High}}   & {\color[HTML]{F8A102} \textbf{Medium}} & {\color[HTML]{FE0000} \textbf{Low}}    \\ \hline
\multicolumn{1}{|l|}{Password Authentication} & {\color[HTML]{009901} \textbf{High}}   & {\color[HTML]{009901} \textbf{High}}   & {\color[HTML]{FE0000} \textbf{Low}}    \\ \hline
\multicolumn{1}{|l|}{Captcha}                  & {\color[HTML]{F8A102} \textbf{Medium}}   & {\color[HTML]{009901} \textbf{High}}   & {\color[HTML]{F8A102} \textbf{Medium}} \\ \hline
\end{tabular}
}
\end{table}

\subsection{Headers Control}

We propose a server-side check on the headers sent with the users' requests. In case a user is accessing the pages from a browser, she would send her specific header \textit{user-agent} \textbf{unclear}. In the same way, the Internet Service Provider (ISP) can set specific \textit{user-agent} on its app requests \textbf{still unclear}. In this way, the remote server can automatically drop all the requests coming from other sources. While this approach limits the access only to specific sources with specific headers, not introducing any friction in the user experience, an attacker can spoof easily the \textit{user-agent} header, making its request appear as an authorized one.

\subsection{In-App Certificate}
It is possible to install specific certificates on the ISP application, to be installed on users' smartphones. The mobile app can sign its requests with those certificates, making the remote server able to identify the request's source. This approach, like the Header Control, does not introduce frictions in the users' experience. However, it is not applicable to browser accesses and, even if harder than spoofing a header, an attacker may be able to extract the certificate and conduct an attack. Another approach in this case would be to expose only limited information (like traffic available, and not the user name) without a specific login. If the remote service exposes only limited information the attack effort may not be justified.
\subsection{SMS Authentication}
When a user wants to access its information, it is possible to ask him to insert its mobile number. The remote system will then check if the mobile number corresponds to the SIM and send a verification code. If the user is able to provide the code, the connection is authenticated and the user can access the service. This approach provides a strong defense, as an attacker must already know the user's mobile number and the verification code received. While this information is available on the smartphone, apps require specific authorization to the user to access it. If the user provides those authorizations, the attacker app is able to perform even worse attacks. This approach introduces a friction for the user, as it has to perform multiple steps adjunctive.
\subsection{Password Authentication}
As in the case of SMS authentication, when the user wants to access her ISP pages, she has to provide some authentication data, which in this case are her username and password. Differently from the SMS based authentication, in this case only the user should know its password, and an attacker is not able to access this information. However the user, in addition to the SMS based authentication steps, has to remember its password.
\subsection{Captcha}
A captcha is a software check which verifies that an action is performed by a real user and not an automatic system. Protecting every access with a captcha makes the access impossible without the interaction with a human. This approach requires a minimum effort for the user, as most recent captcha systems usually require only one click, and protects endpoints from automatic access. On the other hand, a human attacker connected to a smartphone acting as a hotspot is definitely able to overcome such protection directly solving the captcha. However, in this case, or the attacker has already been able to break the WiFi Hotspot protections, like WPA2, or he must be able to obtain the WiFi access key from the smartphone owner.

%
%

\section{Related work}\label{sec:related_work}

The problem of protecting mobile users' privacy is not new. Already in 1999, Jahan et al. addressed security and privacy awareness with a survey for smartphone users \cite{ref1}. Nowadays, with the spread of smartphones, users are monitored more and more, and the need for advanced protection has become evident.

Technically, the mobile operating system producers (\textit{e.g.}, Google with Android and Apple with iOS) introduced an increasing number of {\em permissions} for applications (apps). The users have to explicitly authorize the apps to access particular services or data.  For example, to access the user's position, applications need to ask for multiple permissions. 

In some more restrictive cases, like with iOS, applications are allowed to use these permissions only when the app is foreground. However, Android and iOS are still prey of apps that steal users' information. 
Careless developers often import inside their app third parties libraries without checking their behaviors. Yongzhong \textit{et al.} in \cite{ref5} performed an analysis on more than 150 popular Apps with external libraries, collecting 1909 privacy issues. In some of the issues apps illegitimately connect to remote server and upload data.  And even when the privacy of users is not threatened by flaws in implementation, users' wrong actions may lead to undesired privacy leaks.

Users are often not even aware of how their data may be stolen: Balebako \textit{et al.} \cite{ref2} conducted a test on real participants trying to evaluate the gap between the users' understanding and their real privacy protection. This study reveals how users are not able to clearly understand how to defend themselves. Even most expert users cannot defend themselves in case of privacy flaw in application development.

For this reason, Artz \textit{et al.} \cite{ref3} developed Flowdroid, a tool for taint analysis for android applications able to detect flows which may lead also to privacy leakage inside apps. However, in this case, "features" may become privacy flaws, like the capability of developers of retain users logs and analytics. Indeed,  in \cite{ref4} Liu X. \textit{et al.} demonstrated how public available and widely used analytics libraries for android applications collect private user data.

Differently from the attack presented in our work, collected data is mainly limited to casual events and related to text or positions. In the case presented in our work, exploiting {\vulname} an attacker may collect much more private information belonging to the user.
%
%
%

\section{Conclusions}\label{sec:conclusions}

In the last years, attention to users' privacy is reaching a never seen importance. New patterns, like "Design by Default" or "Privacy by Design", raise the attention to security and privacy to the first position while developing new services. However, users' private data are still far from being protected. 

In this work, we present a vulnerability based on "self-login" which allows collection of personal data from mobile operators. We also show that this vulnerability allows performing active operations, like blocking user's connectivity or stealing credit. We present how several companies are affected by this problem. Although the companies we could test are mainly located in Italy, we want to raise the attention on how the user experience simplification may lead to vulnerabilities. 

In this specific case, it would be enough to introduce a captcha in order to limit programmatic access, to limit the information on self-authenticated pages to the active offers (avoiding private data) and to allow the self-login deactivation (that should be disabled by default in order to follow the "Privacy by default" pattern. Furthermore, we want to highlight how the hot-spot service may be risk-prone if enabled without care. We also present a set of countermeasures that can protect users' privacy.

\begin{IEEEbiography}
[{\includegraphics[width=1in,height=1.25in,clip,keepaspectratio]{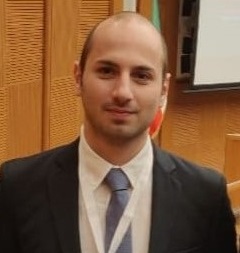}}]{Andrea Coletta}{\,} is a PhD student at the Department of Computer Science at Sapienza University of Rome, Italy. He received his MSc degree in Computer Science form Sapienza University of Rome in Italy. His research interests include wireless networks and mobile network, with particular focus on communication and coordination protocols for aerial drones. Other interests include design and performance evaluation of network protocols for mobile networks. He is student member of IEEE. Contact him at coletta@di.uniroma1.it.
\end{IEEEbiography}

\begin{IEEEbiography}[{\includegraphics[width=1in,height=1.25in,clip,keepaspectratio]{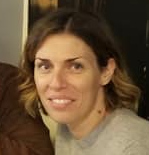}}]{Gaia Maselli}{\,}is an associate professor at the Department of Computer Science at Sapienza
University of Rome, Italy. She holds a Ph.D. in computer science from the University of Pisa, Italy.
Dr. Maselli’s current research interests concern design and implementation aspects of mobile
networks and wireless communications systems, with particular focus on backscattering networks
for the Internet of Things. Other interests include design and performance evaluation of
networking protocols for RFID systems. In the past, Dr. Maselli has contributed to research on
cross-layer design for ad hoc networks.
Dr. Maselli serves as member of the TPC of several international conferences, is associate editor of
Elsevier Computer Communications journal, and serves as reviewer for several journals, such as
IEEE TMC, IEEE TPDS, IEEE TWC, IEEE TON.
Dr. Maselli participated to many European Community research projects such as: CHIRON,
eDIANA, SENDORA, SENSEI, E-sense, WiseNts, MobileMAN. She is member of IEEE. Contact her at maselli@di.uniroma1.it .
\end{IEEEbiography}

\begin{IEEEbiography}[{\includegraphics[width=1in,height=1.25in,clip,keepaspectratio]{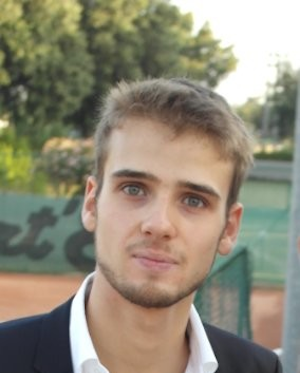}}]{Mauro Piva}{\,} is a PhD Student at the Department of Computer Science. His current research interests include backscattering communication, reinforcement learning for networking, IoT networks, and Software Defined Networking. He is student member of IEEE. Contact him at mauro.piva@uniroma1.it .

\end{IEEEbiography}

\begin{IEEEbiography}[{\includegraphics[width=1in,height=1.25in,clip,keepaspectratio]{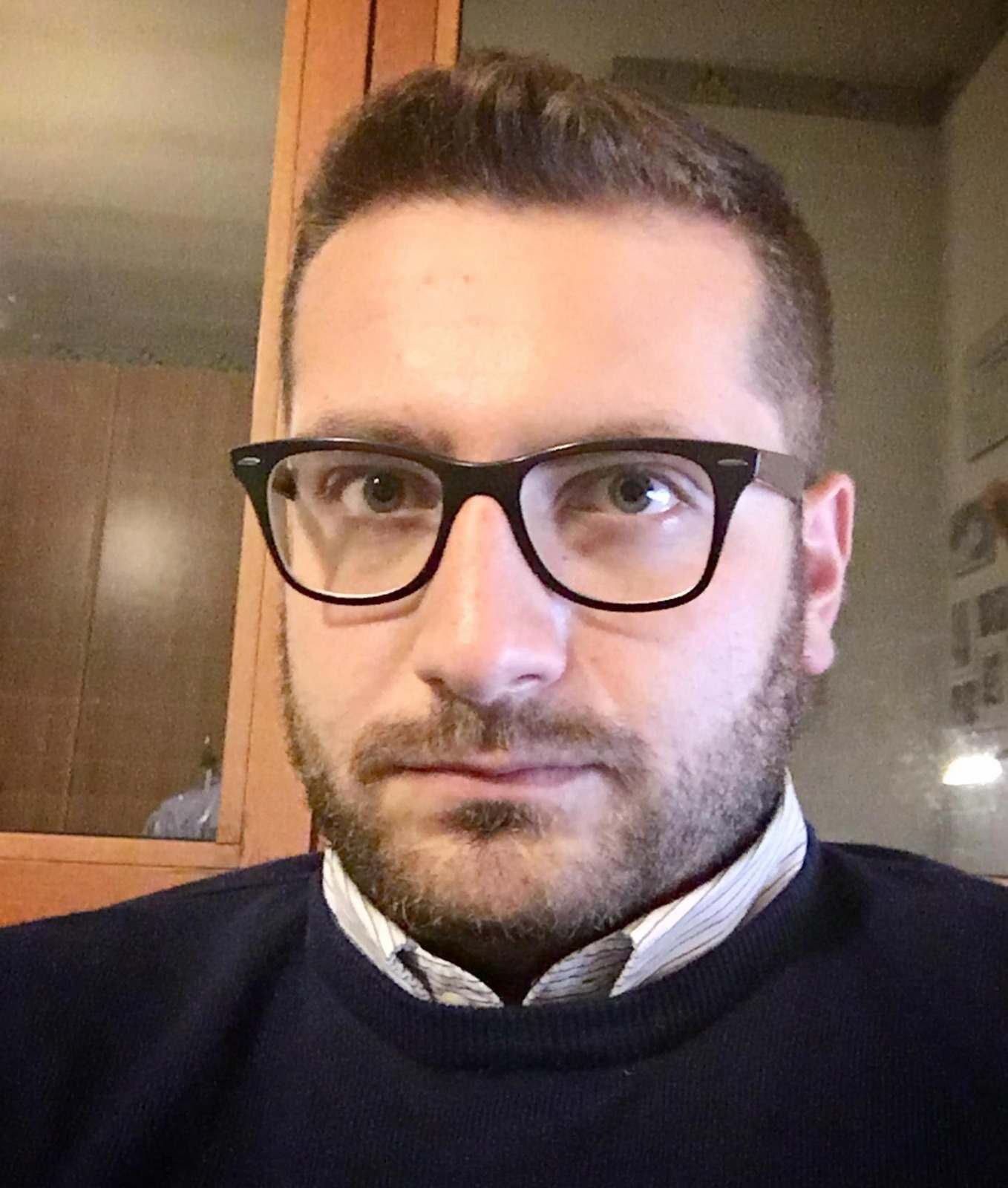}}]{Domenicomichele Silvestri}{\,} received his MSc degree with honors in Computer Science form Sapienza University of Rome, Italy, in 2019 where he is now attending his PhD in Computer Science. His interests include computer networks, architectures, Software Defined Networking, IoT, and drones. Contact him at silvestri.d@di.uniroma1.it .
\end{IEEEbiography}

\begin{IEEEbiography}[{\includegraphics[width=1in,height=1.25in,clip,keepaspectratio]{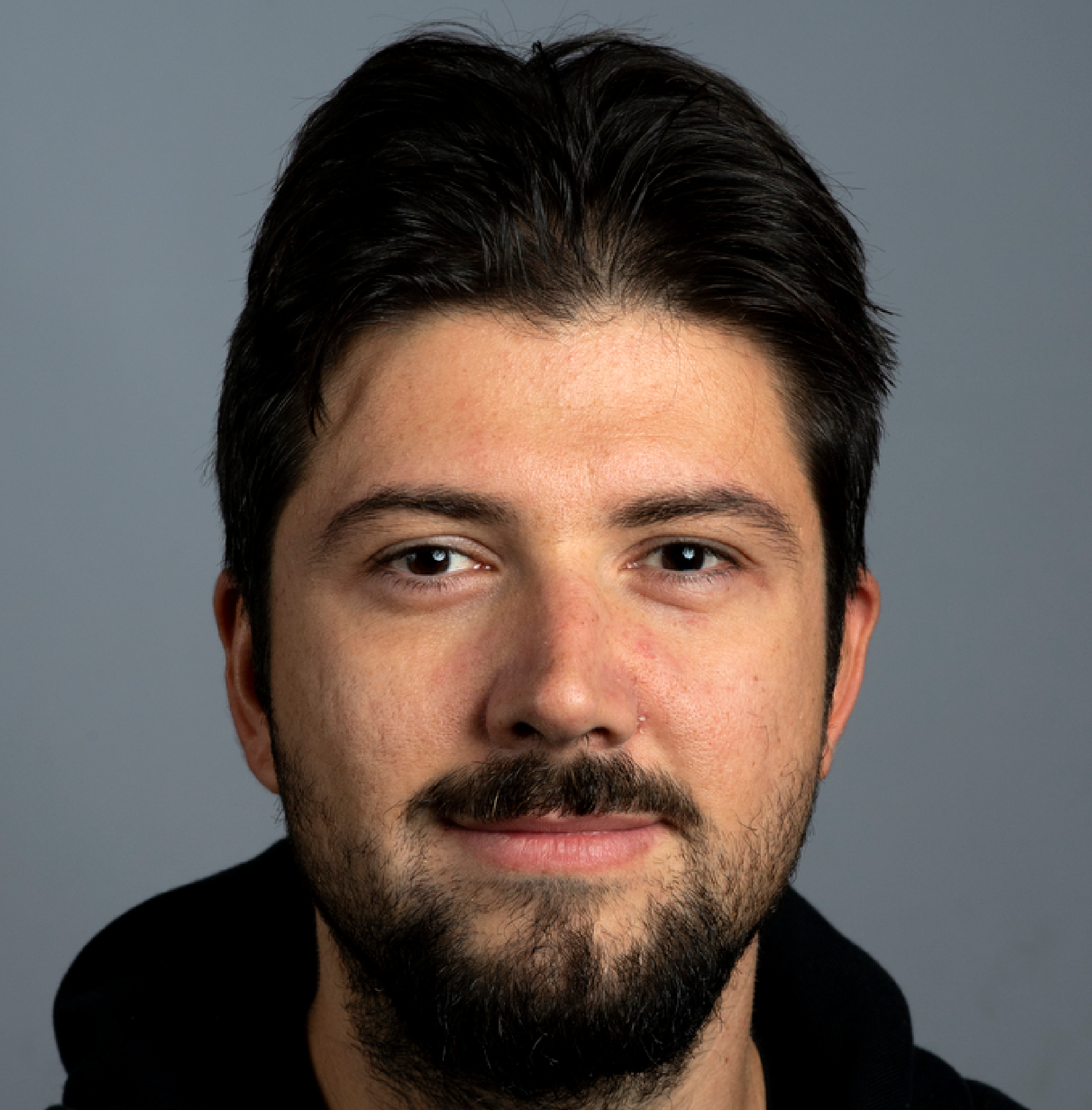}}]{Francesco Restuccia}{\,} (M’16)  is currently an Associate Research Scientist with the Institute for the Wireless Internet of Things, Department of Electrical and Computer Engineering, Northeastern University, Boston, MA, USA. His research interests lie in the modeling, analysis, and experimental evaluation of wireless networked systems. Dr. Restuccia has published over 25 papers in top venues including IEEE INFOCOM, ACM MobiHoc and ACM SenSys, as well as co-authoring 9 pending US patents and 2 book chapters. He regularly serves as a TPC Member for conferences and journals and is a reviewer for several ACM and IEEE conferences and journals. Dr. Restuccia is the recipient of the 2019 Mario Gerla Award for Young Investigators in Computer Science by the Italian Scientists and Scholars of North America Foundation (ISSNAF). He is a Member of the IEEE and the ACM.
\end{IEEEbiography}

\EOD

\end{document}